# Inventory policy for the vaccine of a new pandemic

Aysun Pınarbaşı[*] and Béla Vizvári

Department of Industrial Engineering, Middle East Technical University, Eastern Mediterranean University

[*]corresponding author

**Abstract**

The COVID-19 pandemic showed that stocking and procuring vaccines are different from the inventory holding problems of production. The safety of the supply is more important than cost minimization. The Hungarian inventory model is applied to determine the initial stock if the probability of the non-shortage is given. This initial stock is the function of the procurements. The problem is non-linear as the willingness to be vaccinated is a sigmoid function of the time. The stocking of vaccines is a single-period inventory problem. The cases of three countries are simulated. The countries are Denmark, Hungary, and Mexico. They have different characteristics of willingness to be vaccinated. The numerical results of the simulation show that the prescribed probability of non-shortage can be achieved.

## 1. Introduction

COVID-19 is the first pandemic in the history of mankind detailed data are available on the number of cases and the used vaccines in every country of the world (Our World in Data, n.d.). This fact gives the unique opportunity to determine the theoretical bases of the logistics of future pandemics. The COVID-19 pandemic raised many questions which can be solved only by scientific methods. Many answers can be used only in the case of future pandemics. The reason is that the correct understanding of the problem and the elaboration of the solution takes time. The economic effects of the pandemic belong to that category. In this paper, an inventory strategy is elaborated for the procurement of vaccines from the point of view of individual states. The strategy finds a balance between safety and costs.

The rich individual countries purchased high initial stocks to make sure that there would be no shortage of the vaccine. The stocks could not be used in some countries as the willingness to be vaccinated is limited in every country. Later, the countries presented superfluous quantities of vaccines to poor countries or sold them at a low price. This behavior is bad for every agent of the market. The countries that purchased too many vaccines lost money. The pharmaceutical companies could not meet all the needs. They may lose markets in the future. The countries that did not get the vaccine in time had a more serious pandemic. Many people died unnecessarily. Finally, the world as a single closed system had and still has the pandemic for a longer period because of the missing vaccines as people from low-income countries can infect people from high-income countries.

The Hungarian inventory model is applied to obtain optimal inventory policy for vaccines. This model keeps both the probability of shortage and the initial stock low in general. Most of the inventory models assume that the time horizon is infinitely long. The Hungarian inventory model can be applied both in a multi-period and a single-period environment. The latter one is important concerning the pandemic. In section 2, the stocking process of vaccines and some lessons of the COVID-19 pandemic are introduced for selected countries. In section 3, the single period inventory problem; In section 4, The Hungarian inventory problem has been discussed. Section 5 mentioned basic mathematical facts about the current problem. Section 6 showed, how to apply the theory to the vaccination problem. Section 7 discusses the numerical results of a simulation. In section 8, business and safety considerations are discussed. The paper is finished with conclusions.

**2. The stocking process of vaccines and some lessons of the COVID-19 pandemic**

Few pharmaceutical companies are able to produce a certain vaccine. Production is a long-term process. Another problem is that if the production process stops for any reason, then it is time-consuming to restart the process. One important lesson of the COVID-19 pandemic was that a company must satisfy the demands of many countries if a new disease occurs. There was a shortage of COVID-19 vaccines ("Unacceptable" Vaccine Delays Cause Frustration across the European Union | World News | Sky News, 2021). This fact cannot be explained only by the high demand. This type of vaccine is a new product causing some technical difficulties. However, the shortage can occur even from regular vaccines such as diphtheria(U.N. Reports Shortage Of Diphtheria Vaccine - The New York Times, 1994) even in countries like the USA (Diphtheria-Tetanus-Pertussis Vaccine Shortage -- United States, 1984) and European Union (April - Shortage of Essential Diphtheria Treatment Drugs Needs International Action, Experts Warn - University of Exeter, 2017).

The distribution of vaccines consists of at least two phases. A country buys a large quantity. The vaccine generally needs cooling. The purchased quantity is distributed inside the country in the second step. There are diseases such that one vaccine is enough per person. Most countries suggest not more than four vaccines per person in case of COVID-19. Thus, the size of the concerned population gives an upper bound of the demand of the country as the willingness to be vaccinated is less than 100 percent and some people may not be vaccinated because of medical reasons.

Another important factor is that the willingness is changing over time. This phenomenon was observed in the case of COVID-19. The pandemic has several waves. See the four waves of Hungary in Figure 1. There was no vaccine at the beginning. The vaccination started much later than the pandemic. Thus, the profile of vaccination is different. For example, Denmark had two waves of the pandemic. First, there was a weak one and it was followed by a strong wave. However, the two waves of vaccination had approximately the same strength (see Figures 2 and 3). The COVID-19 pandemic has some special properties. The total number of vaccinations could exceed the population of the country as a person needs more than one vaccine. Similarly, the number of cases is higher than the number of people who suffer from the disease as a person could get it several times. Concerning COVID-19, the total demand of a country is bounded as no more than four vaccines are suggested to a person. Small children and persons having certain diseases cannot be vaccinated. Moreover, there are people against vaccination, unfortunately. Thus, the upper bound is less than a certain multiple of the total population of the country. The countries showed similar behavior during the COVID-19 pandemic. However, they had individual parameters depending on the size, location, and cultural background of the country. It is supposed that the behaviors of the countries will be similar in the case of future pandemics. However, some parameters, like total demand, may depend on the properties of the pandemic and future technology.

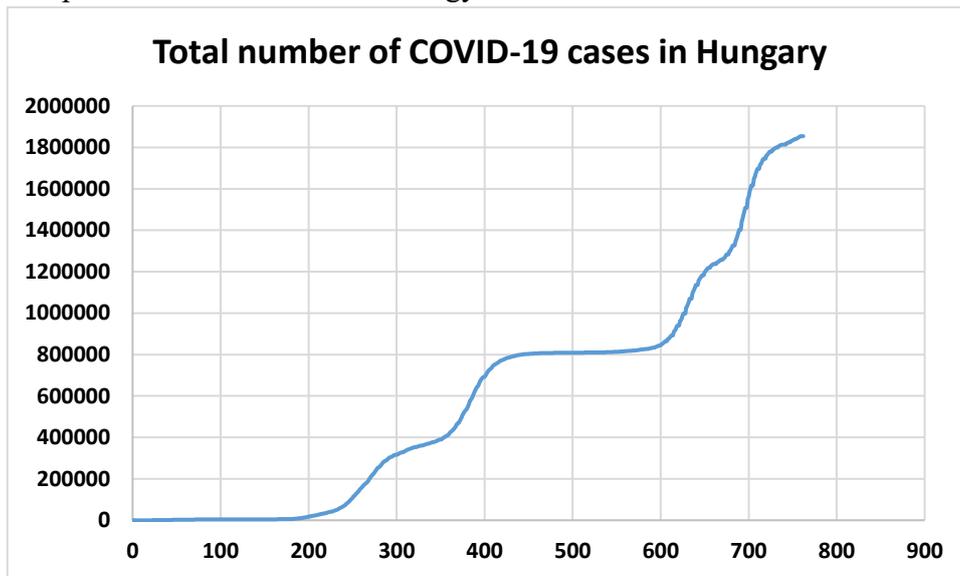

**Figure 1.** The total number of COVID-19 cases in Hungary. The pandemic started in the country on 2020.03.04. The time interval o the waves in days are as follows: 1-st [1,332], 2-nd [333,500], 3-rd [501,685], and 4-th [686,762].

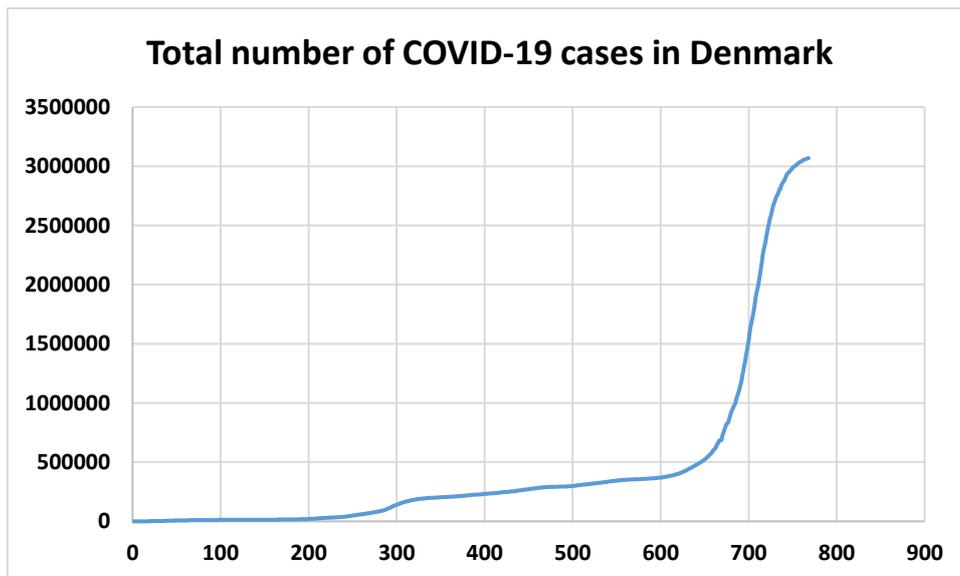

**Figure 2.** The waves of the pandemic in Denmark. The inflection point of the first one is around day 300. The first pandemic case was reported on 2020.02.27.

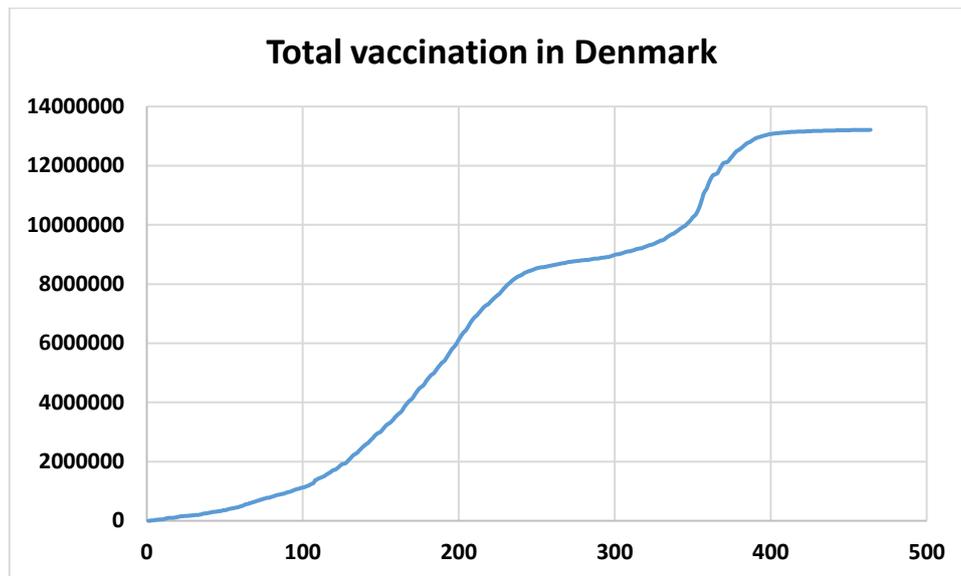

**Figure 3.** The two waves of vaccination in Denmark. Vaccination started at 2020.12.27 not considering very few cases in December of 2020.

**3. Single period inventory problem**
The first ever paper on an inventory problem is (Harris, 1990). It determines the Economic Order Quantity (EOQ) in a deterministic case such that there are infinitely many time-periods and all time-periods are completely identical. Still, the majority of inventory problems have an infinite time horizon. An important exception is the newsboy problem (Chen *et al.*, 2016). The decision of the sock level is made for a single period in this case.

Is vaccination a single-period or a multi-period inventory problem? There is no unique answer to this question as there are many different vaccination problems. Every country applies a routine immunization profile which is a standardized schedule of giving the vaccine to main children and preventing the outbreak of certain pandemics such as diphtheria, measles, smallpox, etc. The realization of the profile is repeated every year. Thus, the inventory problem of routine immunization is having an infinite time horizon.

The case of COVID-19 is quite different. It is a new pandemic. Thus, there is no routine immunization against it. The situation was even worse at the beginning as no vaccine existed against this disease. The general idea was that if a vaccine is developed and the population is vaccinated, then the pandemic will stop. If the assumption is correct, then it is a single-period problem. The period is the duration of the pandemic. The same can be expected in the future in the case of still-unknown pandemics. The three main steps will be as follows: (a) developing the vaccine, (b) producing the vaccine in large quantities, and (c) vaccinating the population.
The above-mentioned assumption is applied in this paper. Thus, there is only one period.

The solution to the inventory problem starts with the determination of the demand. The demand of a country depends on two main characteristics of the country which are the size of the population and the shape of the willingness to be vaccinated. The further steps of the solution are the determination of the number of purchases including the initial one and the purchased quantities. This paper provides a solution to the mentioned decision structure taking into account the stochastic nature of the market.

## 4. The Hungarian inventory model

The Hungarian inventory model is another example of a single-period inventory problem [8]. The model was developed in the early 1960s (Prékopa, 2006). It is also called the Prékopa-Ziermann model. More precisely, it can be used both in single-period and multi-period cases. It is discussed as a multi-period problem first. The raw material is consumed by the production in the model. Its procurement is uncertain. The safety of the production, i.e. avoiding unwanted interruption of the production, is more important than the inventory cost. The main question of the Hungarian inventory model is how large should be the initial stock such that the probability of shortage is not greater than an *a priori* given threshold. In the simplest version of the model, the uncertainty is in the delivery times. The uncertainty is modeled by a stochastic approach.

The basic assumptions and the mathematical formalism of the model are as follows:
- The events take place in the time interval $[0, T]$ where $T > 0$.
- The initial stock is $M$. It is unknown and is obtained from the optimal solution of the model.
- The demand is deterministic, and its value is $D$ in the time interval.
- There are $n$ deliveries in the time interval where $n$ is a positive integer. This number does not contain the delivery of the initial stock.
- Each delivery transports the same quantity which is $D/n$.
- The deliveries take place in uniformly distributed and independent times, say at $t_1, t_2, \ldots, t_n$. These values are uniformly distributed and independent on the interval $[0, T]$, thus the index order may be different from their increasing order. The history of the deliveries, i.e. when they happen in time order, is given by their order statistics $t_1^\square < t_2^\square < \cdots < t_n^\square$ where $\{t_1, t_2, \ldots, t_n\}$ and $\{t_1^\square, t_2^\square, \ldots, t_n^\square\}$ run over the same set of real numbers
- There is a probability value $p(0 < p < 1)$ given. It is the required probability of the non-shortage.

Two further functions need special attention. Both are the function of time $t$, where $0 \leq t \leq T$. The first one is the total demand until time $t$. It is denoted by $F(t)$. It was supposed in the original Hungarian inventory model to be linear, like $F(t) = ct$ where $c$ is a positive constant (Prékopa, 2006). However, it is not linear in the case of vaccination as Figure 3 shows. Concerning the second function, let $W(n, t)$ be the total available quantity of vaccine until time $t$. It is a random function as the delivery times are random. Its value is $M$ plus $D/n$ multiplied by the number of deliveries until time $t$. Its mathematical formula is as follows:

$$W(n, t) = M + \frac{kD}{n}, where\, t_k^\square \leq t < t_{k+1}^\square. \tag{1}$$

The default values $t_0^\square = 0$ and $t_{n+1}^\square = +\infty$ are applied. Formula (1) concerns to the multi-period case. It is important to note that the term $M$ is deterministic and constant on the time interval $[0, T]$. The term $\frac{kD}{n}$ is constant; however, the interval where it is added depends on random factors. The material balance in the $[0, T]$ interval is, that the initial stock level is $M$ and the delivered total quantity is $n(\frac{D}{n}) = D$, i.e. the total available quantity is $M + D$. On the other hand, the used quantity is $D$. Thus, at the end of the interval the remaining quantity is $(M + D) - D = M$. It is the initial stock of the next time period. Thus, the system is repeating itself. However, if there is only one period as in the case of the pandemic, then the stock level is to be decreased to the end of the period to 0. Thus, the initial stock is $M$ again and the total quantity of the procurement is $D - M$. Hence, the delivered quantity is $D/n$ if at least this quantity is missing from the total demand $D$. If less but still positive value is missing, then the purchased quantity is the missing quantity and finally, if the available quantity achieved $D$, then the purchased quantity is 0. Let $Y(n,k)$ be the quantity which is purchased at the $k$-th delivery. Then

$$Y(n,k) = \begin{cases} D/n & If D \geq M + kD/n \\ D - M - (k-1)D/n & If M + \frac{(k-1)D}{n} < D \leq m + kD/n \\ 0 & If M + (k-1)D/n \geq D \end{cases} \quad (2)$$

It is easy to see that (2) is equivalent to

$$Y(n,k) = min\{D/n, max\{0, D - M - (k-1)D/n\}\}. \quad (3)$$

The mathematical condition that there is no shortage at a moment $t$ is that the total arrived quantity is at least as large as the demand, i.e.

$$W(n,t) \geq F(t).$$

Similarly, the condition that there is never shortage is that the same inequality holds in every moment, i.e.

$$\forall t \in [0,T]: W(n,t) \geq F(t). \quad (4)$$

The satisfaction of condition (4) is a random event. Its probability depends on the value of the initial stock $M$. The probability is an increasing function of $M$. The Hungarian inventory model claims to determine the minimal value of $M$ such that this probability is at least $p$. The mathematical formulation of the model is as follows:

$$min_M. Prob(\forall t \in [0,T]: W(n,t) \geq F(t)) \geq p. \quad (5)$$

Thus, the optimal solution of the model gives the minimal initial stock that the vaccination process can be carried out without interruption with probability $p$.

Formula (1) allows that the total available quantity, that is the initial stock and the total delivered quantity is higher than the demand in a period. It is not necessary to purchase that much in a single-period case. No shortage is created if at time $t_k^{\square}$ quantity $Y(n,t)$ is delivered. Hence, the total available quantity is:

$$X(n,t) = M + \sum_{l=1}^{k} Y(n,l), where \, t_k^{\square} \leq t < t_{k+1}^{\square}. \quad (6)$$

$X(n,t)$ is the single period version of $W(n,t)$.

## 5. Basic mathematical facts from [9]

The mathematical bases of [8] can be found in (Birnbaum and Tingey, 1951). This paper introduced the *upper confidence contour*. Similar results can be found in (Bernstein, 1946). The same notations are used as in the previous section. The staircase function $W(n,t)$ is closely related to the *empirical distribution function* of the total delivered quantity. Let $Z(n,t)$ be the empirical distribution function, i.e.

$$Z(n,t) = \frac{k}{n}, if \, t_k^{\square} \leq t < t_{k+1}^{\square}, k = 1, \dots, n; Z(n,t) = 0, if \, t \leq 0; Z(n,t) = 1, if \, t \geq T. \quad (7)$$

Hence, $W(n,t) = M + Z(n,t)D$, i.e., the function $W(n,t)$ is obtained if $Z(n,t)$ is multiplied by $D$ and is shifted up by $M$.

The upper confidence contour is a function which majorizes the empirical distribution function. Let $\varepsilon > 0$ be a real number. The sample size is still $n$. The upper confidence contour is defined by the equation:

$$Z(n, \varepsilon, t) = min\{Z(n, t) + \varepsilon, 1\}. \tag{8}$$

There is a simple homogeneous linear relation of the single period available quantity and the upper confidence contour. Assume that $M = \varepsilon D$. Notice that it implies that:

$$X(n, t) = Z(n, t)D. \tag{9}$$

(Birnbaum and Tingey, 1951) introduced a distribution-independent quantity which depends on two factors as follows: (i) the sample size $n$, and (ii) $\varepsilon$. It is the probability that upper confidence contour is above the distribution function of the sample. It is denoted by $P(n, \varepsilon)$ and its mathematical formula is:

$$P(n, \varepsilon) = Prob\left(\forall F: \left(\forall t: Z(n, \varepsilon, t) \geq F(t)\right)\right).$$

The main result of (Birnbaum and Tingey, 1951) is the value of $P(n, \varepsilon)$.

**Theorem** (Birnbaum and Tingey, 1951)**:** For all positive integer $n$ and real number $0 < \varepsilon \leq 1$

$$P(n, \varepsilon) = 1 - \varepsilon \sum_{j=0}^{\lfloor n(1-\varepsilon) \rfloor} \binom{n}{j} \left(1 - \varepsilon - \frac{j}{n}\right)^{n-j} \left(\varepsilon + \frac{j}{n}\right)^{j-1}. \tag{10}$$

In what follows, the solution of the equation in $\varepsilon$ that (10) is equal to $p$ is denoted by $\varepsilon(n, p)$, i.e.

$$P\bigl(n, \varepsilon(n, p)\bigr) = p.$$

**6. How to apply the theory to the vaccination problem?**

The application of the Hungarian inventory model is discussed for a single-wave pandemic.

The total willingness to be vaccinated as the function of time is not linear in time (Pınarbaşı and Vizvári, 2022). It can be approximated by sigmoid functions including even the *arctan* function. The reason is mentioned in section 2. It is that each person gets only a restricted number of vaccines. Thus, the total number of used vaccines as a function of time is increasing and has a horizontal asymptote.

The necessary condition for the applicability of the results discussed in the previous section is that the delivery times must be drawn from a distribution having a proper cumulative distribution function. The *arctan* function is transformed into a cumulative distribution function as:

$$\frac{arctan\,(t)}{\pi} + \frac{1}{2}.$$

This function is the cumulative distribution function of the Cauchy distribution. However, it must be fitted to the function of the particular country by some parameters. The general form of the fitted function is:

$$\frac{a\,arctan\,(b(t-c))}{\pi} + d.$$

The meanings of the parameters are as follows: $a$ is the parameter of the vertical stretching, $b$ is the horizontal stretch of the function, and the inflection point is $c$, the vertical shift if $d$.

The daily data of the total number of vaccinated people in countries is available in (Our World in Data, n.d.). When a particular country is investigated, then a new time series is to be created. The letter $t$ indicating the time means the number of the day. Similarly, the last day of the observations is $T$ and the first day is 1. The new time series is:

$$\frac{the\ total\ number\ of\ vaccines\ used\ until\ day\ t}{the\ total\ number\ of\ vaccines\ used\ until\ day\ T}.$$

This value is between 0 and 1. Furthermore, it is monotone increasing. Any decrease indicates an error in the data collection. It happened in the case of France (Our World in Data, n.d.). The delivery times must be drawn from the distribution fitted to the generated time series.

Hence, the necessary initial stock is:

$$\varepsilon(n,p) \times the\ total\ number\ of\ used\ vaccines\ until\ day\ T. \tag{11}$$

However, it is obvious that the "*the total number of vaccines used until day T*" is *a posteriori* information. It can be estimated by the total number of vaccines used in previous pandemics. The size of the population is an upper bound of it as it is supposed that there is only one wave of the pandemic. The number of small children and the number of people suffering from special diseases can be deduced from the size of the total population. All these considerations give an estimation of the total number of vaccinated people giving the initial stock instead of (11) one vaccine per person is enough:

$$\varepsilon(n,p) \times the\ estimated\ number\ of\ vaccinated\ people. \tag{12}$$

The results can be checked by simulation.

It is important to remark that (Pınarbaşı and Vizvári, 2022) shows that even during the pandemic it is possible to obtain an estimation of good accuracy of used vaccines by curve fitting. Estimation obtained only 64 days after the outbreak of the pandemic has good quality. The error of the estimation obtained 94 days after the outbreak is small.

**7. Results of numerical simulation**

(Birnbaum and Tingey, 1951) also determined the value of $\varepsilon(n,p)$ for $n = 5, 8, 10, 20, 40, 50, p = 0.9, 0.95, 0.99, 0.999$. The values for $p = 0.9$ are reconstructed in Table 1. The quantity of each procurement is the same. Assume that the achieved total vaccination percentage is known *a priori*. Then the quantity of procurement is:

$$\frac{total\ population \times vaccination\ percentage}{the\ number\ of\ procurements}.$$

Without loss of generality, it is assumed in the simulation that the size of the total population of a country is 1. The assumption is based on the homogeneous linear relation (9), *i.e.* it is justified by the fact that both the initial stock and the procurement quantity are proportional to the total population.

**Table 1.** The relation of the number of procurements and the necessary initial stock to achieve 90 percent probability of non-shortage.

| Number of procurements | Relative initial stock |
|---|---|
| 5 | 0.4470 |
| 8 | 0.3583 |
| 10 | 0.3226 |
| 20 | 0.23155 |
| 40 | 0.16547 |
| 50 | 0.14840 |

It is recalled that the function $P(n,\varepsilon)$ is distribution independent. Thus, if the initial stock is exactly the value given in Table 1, then the achieved probability for a particular distribution can be higher than 0.9.

Three countries are involved in the simulation. They are Denmark, Hungary, and Mexico. The required probability of non-shortage is 0.9. The numbers of procurements are 5, 8, and 10. The higher number of procurements is not realistic. The three countries were selected because of their different culture and size.

The simulation is carried out as follows: The vaccination speeds of the countries are approximated by the fitted *arctan* function. 44.7, 35.83, and 32.26 percentages were the initial stock for the number of procurements 5, 8, and 10, respectively. The quantities of the procurements are 20, 12.5, and 10 percent, respectively. 10,000 scenarios are generated randomly in each of the nine cases. Each scenario consists of as many delivery days as the number of procurements. Each scenario is checked such that if the procurement quantities arrive on the days of the scenario, then whether or not shortage exists at any time. The ratio of the non-shortage scenarios is the simulated value of the probability.

**Table 2.** The case of Denmark is simulated for 5, 8, and 10 procurements. The achieved probabilities of non-shortage in case of different procurement quantities are provided.

| Achieved percentage | Regular procurement in percentage of the total population | Probability | Achieved percentage | Regular procurement in percentage of the total population | Probability | Achieved percentage | Regular procurement in percentage of the total population | Probability |
|---|---|---|---|---|---|---|---|---|
|  | 0,188 | 0,889 |  | 0,113 | 0,8673 |  | 0,088 | 0,842 |
| 100,00% | 0,189 | 0,8911 | 100,00% | 0,114 | 0,8724 | 100,00% | 0,089 | 0,8479 |
|  | 0,19 | 0,8911 |  | 0,115 | 0,8753 |  | 0,09 | 0,856 |
|  | 0,191 | 0,8935 |  | 0,116 | 0,8791 |  | 0,091 | 0,8628 |
| Number of procurements | 0,192 | 0,8935 | Number of procurements | 0,117 | 0,8863 | Number of procurements | 0,092 | 0,8688 |
|  | 0,193 | 0,8973 |  | 0,118 | 0,8876 |  | 0,093 | 0,8793 |
| 5 | 0,194 | 0,8995 | 8 | 0,119 | 0,8923 | 10 | 0,094 | 0,8816 |
|  | 0,195 | 0,8995 |  | 0,12 | 0,8924 |  | 0,095 | 0,8877 |
|  | 0,196 | 0,8995 |  | 0,121 | 0,8974 |  | 0,096 | 0,8923 |
| Initial stock in percentage of total population | 0,197 | 0,901 | Initial stock in percentage of total population | 0,122 | 0,8989 | Initial stock in percentage of total population | 0,097 | 0,896 |
|  | 0,198 | 0,901 |  | 0,123 | 0,9001 |  | 0,098 | 0,9022 |
|  | 0,199 | 0,903 |  | 0,124 | 0,9034 |  | 0,099 | 0,9065 |
|  | 0,2 | 0,906 |  | 0,125 | 0,9062 |  | 0,1 | 0,9096 |
| 0.447 | 0,201 | 0,906 | 0.3583 | 0,126 | 0,9086 | 0.3226 | 0,101 | 0,912 |
|  | 0,202 | 0,9075 |  | 0,127 | 0,9095 |  | 0,102 | 0,9144 |
|  | 0,203 | 0,9075 |  | 0,128 | 0,9115 |  | 0,103 | 0,9185 |
|  | 0,204 | 0,9075 |  | 0,129 | 0,9128 |  | 0,104 | 0,9206 |
|  | 0,205 | 0,909 |  | 0,13 | 0,9136 |  | 0,105 | 0,923 |
|  | 0,206 | 0,909 |  | 0,131 | 0,9172 |  | 0,106 | 0,9254 |
|  | 0,207 | 0,9102 |  | 0,132 | 0,9178 |  | 0,107 | 0,9296 |

| | 0,208 | 0,9131 | | 0,133 | 0,9209 | | 0,108 | 0,9301 |

**Table 3.** The case of Hungary is simulated for 5, 8, and 10 procurements. The achieved probabilities of non-shortage in case of different procurement quantities are provided.

| Achieved percentage | Regular procurement in percentage of the total population | Probability | Achieved percentage | Regular procurement in percentage of the total population | Probability | Achieved percentage | Regular procurement in percentage of the total population | Probability |
|---|---|---|---|---|---|---|---|---|
| | 0,188 | 0,8992 | | 0,113 | 0,8701 | | 0,088 | 0,8362 |
| **100,00%** | 0,189 | 0,8992 | **100,00%** | 0,114 | 0,8719 | **100,00%** | 0,089 | 0,8514 |
| | 0,19 | 0,9014 | | 0,115 | 0,879 | | 0,09 | 0,857 |
| **Number of procurements** | 0,191 | 0,9014 | **Number of procurements** | 0,116 | 0,879 | **Number of procurements** | 0,091 | 0,8647 |
| | 0,192 | 0,9014 | | 0,117 | 0,8801 | | 0,092 | 0,8718 |
| | 0,193 | 0,9014 | | 0,118 | 0,8873 | | 0,093 | 0,8774 |
| **5** | 0,194 | 0,9014 | **8** | 0,119 | 0,8956 | **10** | 0,094 | 0,8774 |
| | 0,195 | 0,9031 | | 0,12 | 0,8972 | | 0,095 | 0,8919 |
| | 0,196 | 0,9031 | | 0,121 | 0,8976 | | 0,096 | 0,8931 |
| **Initial stock in percentage of total population** | 0,197 | 0,9072 | **Initial stock in percentage of total population** | 0,122 | 0,9029 | **Initial stock in percentage of total population** | 0,097 | 0,8968 |
| | 0,198 | 0,9072 | | 0,123 | 0,9029 | | 0,098 | 0,9022 |
| | 0,199 | 0,9089 | | 0,124 | 0,9037 | | 0,099 | 0,9056 |
| | 0,2 | 0,9089 | | 0,125 | 0,9066 | | 0,1 | 0,9063 |
| **0,447** | 0,201 | 0,9089 | **0,3583** | 0,126 | 0,912 | **0,3226** | 0,101 | 0,9148 |
| | 0,202 | 0,9089 | | 0,127 | 0,912 | | 0,102 | 0,9153 |
| | 0,203 | 0,9089 | | 0,128 | 0,912 | | 0,103 | 0,9194 |
| | 0,204 | 0,911 | | 0,129 | 0,913 | | 0,104 | 0,9221 |
| | 0,205 | 0,911 | | 0,13 | 0,9181 | | 0,105 | 0,9236 |
| | 0,206 | 0,911 | | 0,131 | 0,9224 | | 0,106 | 0,925 |
| | 0,207 | 0,911 | | 0,132 | 0,9224 | | 0,107 | 0,9299 |
| | 0,208 | 0,9174 | | 0,133 | 0,9254 | | 0,108 | 0,9306 |

**Table 4.** The case of Mexico is simulated for 5, 8, and 10 procurements. The achieved probabilities of non-shortage in case of different procurement quantities are provided.

| Achieved percentage | Regular procurement in percentage of the total population | Probability | Achieved percentage | Regular procurement in percentage of the total population | Probability | Achieved percentage | Regular procurement in percentage of the total population | Probability |
|---|---|---|---|---|---|---|---|---|
| | 0,188 | 0,8902 | | 0,113 | 0,86 | | 0,088 | 0,8332 |
| **100,00%** | 0,189 | 0,8922 | **100,00%** | 0,114 | 0,8646 | **100,00%** | 0,089 | 0,8422 |
| | 0,19 | 0,893 | | 0,115 | 0,8698 | | 0,09 | 0,8508 |
| **Number of procurements** | 0,191 | 0,893 | **Number of procurements** | 0,116 | 0,8725 | **Number of procurements** | 0,091 | 0,8574 |
| | 0,192 | 0,8945 | | 0,117 | 0,8766 | | 0,092 | 0,8664 |
| | 0,193 | 0,8947 | | 0,118 | 0,8811 | | 0,093 | 0,8715 |
| **5** | 0,194 | 0,8954 | **8** | 0,119 | 0,8853 | **10** | 0,094 | 0,8754 |
| | 0,195 | 0,8961 | | 0,12 | 0,8894 | | 0,095 | 0,88 |
| | 0,196 | 0,8975 | | 0,121 | 0,891 | | 0,096 | 0,8867 |
| **Initial stock in percentage of total population** | 0,197 | 0,8979 | **Initial stock in percentage of total population** | 0,122 | 0,894 | **Initial stock in percentage of total population** | 0,097 | 0,8901 |
| | 0,198 | 0,8984 | | 0,123 | 0,896 | | 0,098 | 0,8931 |
| | 0,199 | 0,9009 | | 0,124 | 0,8982 | | 0,099 | 0,8996 |
| | 0,2 | 0,9014 | | 0,125 | 0,9011 | | 0,1 | 0,9029 |
| **0,447** | 0,201 | 0,9022 | **0,3583** | 0,126 | 0,9033 | **0,3226** | 0,101 | 0,9061 |
| | 0,202 | 0,9036 | | 0,127 | 0,9059 | | 0,102 | 0,9088 |
| | 0,203 | 0,9039 | | 0,128 | 0,9069 | | 0,103 | 0,9133 |
| | 0,204 | 0,9041 | | 0,129 | 0,9096 | | 0,104 | 0,9167 |

| | 0,205 | | | 0,13 | 0,9115 | | 0,105 | 0,9179 |
|---|---|---|---|---|---|---|---|---|
| | 0,206 | 0,9068 | | 0,131 | 0,9136 | | 0,106 | 0,9213 |
| | 0,207 | 0,907 | | 0,132 | 0,9162 | | 0,107 | 0,923 |
| | 0,208 | 0,9077 | | 0,133 | 0,9179 | | 0,108 | 0,9259 |

## 8. Business and safety considerations

Table 1 concerns all types of products not only vaccines. The values for probability levels other than 0.9 are similar to those in Table 1. More precisely, the higher the probability of the non-shortage, the higher the initial stock level is. On the other hand, the initial stock level is a decreasing function of the number of procurements. If the cost is the most important factor, then a good compromise is to be found depending on the order cost.

Safety is the most important factor in the case of vaccines. If a country may purchase a high initial stock, then this country is safe with a high probability. However, there were cases in the COVID-19 pandemic such that a country could not use all the initial stock. The country suffered financial loss even in the case of a part of the superfluous stock has been sold.

Another drawback that a country has too high initial stock is that other countries can have a problem with procurement as not enough vaccines remain on the market. If other countries are unable to vaccinate their citizens, then the infection will continue to spread there, and this will re-infect, to a greater or lesser extent, the citizens of countries that have come in safety. Therefore, at the international level, it is desirable to develop a system in which each country is guaranteed a lot of deliveries evenly distributed in time.

## 9. Conclusions

The stocking procedure of vaccines in an individual country is investigated. The outbreak of a pandemic is assumed. Vaccination can stop the pandemic. Thus, there is a sudden high demand for the vaccine which causes uncertainty in deliveries. The Hungarian inventory model is a proper tool to determine the optimal inventory policy of an individual country. The results of the model are analyzed on the data of the COVID-19 pandemic. As both the spreading and vaccination of COVID-19 have special non-linear properties, a non-linear version of the Hungarian inventory model is applied. The outcomes of the model were checked by simulation of three countries as follows: Denmark, Hungary, and Mexico. The results of the simulation show that an even higher probability of non-shortage can be achieved than what the model promises.